\documentclass[%
 reprint,
 superscriptaddress,
 amsmath,amssymb,
 aps,
]{revtex4-2}

\usepackage{graphicx}
\usepackage{dcolumn}
\usepackage{bm}

\begin{document}

\preprint{APS/123-QED}

\title{A Temporal Retrieval Method for Modulated Electron Bunches via Adaptive Kernel Reconstruction}

\author{Zixiao Guo}
\affiliation{%
 State Key Laboratory of Ultra-intense Laser Science and Technology, Shanghai Institute of Optics and Fine Mechanics (SIOM), Chinese Academy of Sciences (CAS), Shanghai 201800, China
}
\affiliation{%
 Center of Materials Science and Optoelectronics Engineering, University of Chinese Academy of Sciences (UCAS), Beijing 100049, China
}
\author{Ke Feng}
 \email{fengke@siom.ac.cn}
\affiliation{%
 State Key Laboratory of Ultra-intense Laser Science and Technology, Shanghai Institute of Optics and Fine Mechanics (SIOM), Chinese Academy of Sciences (CAS), Shanghai 201800, China
}
\author{Zhiheng Lou}
\affiliation{%
 State Key Laboratory of Ultra-intense Laser Science and Technology, Shanghai Institute of Optics and Fine Mechanics (SIOM), Chinese Academy of Sciences (CAS), Shanghai 201800, China
}
\author{Guiyao Wang}
\affiliation{%
 State Key Laboratory of Ultra-intense Laser Science and Technology, Shanghai Institute of Optics and Fine Mechanics (SIOM), Chinese Academy of Sciences (CAS), Shanghai 201800, China
}
\author{Wentao Wang}
 \email{wwt1980@siom.ac.cn}
\affiliation{%
 State Key Laboratory of Ultra-intense Laser Science and Technology, Shanghai Institute of Optics and Fine Mechanics (SIOM), Chinese Academy of Sciences (CAS), Shanghai 201800, China
}
\author{Ruxin Li}
 \email{ruxinli@mail.siom.ac.cn}
\affiliation{%
 State Key Laboratory of Ultra-intense Laser Science and Technology, Shanghai Institute of Optics and Fine Mechanics (SIOM), Chinese Academy of Sciences (CAS), Shanghai 201800, China
}
\affiliation{%
 School of Physical Science and Technology, ShanghaiTech University, Shanghai 201210, China
}

%




\date{\today}

\begin{abstract}
Femtosecond electron beams with complex modulation play a crucial role in 
applications such as X-ray Free Electron Lasers (XFELs) and plasma wakefield 
accelerators. However, diagnostics for the electron beam current profile 
still face challenges with complex structure. In this letter, 
we propose a novel temporal retrieval algorithm for the coherent 
transition radiation (CTR) diagnostics of complex modulated electron beams. 
Starting from the time-frequency analysis of the electron bunch train, 
the algorithm separates and reconstructs the high- and low-frequency components. 
A temporal kernel was derived from the inverse sampling of the measured spectrum to 
construct the high-frequency component, while the low-frequency 
envelope was composed of several basis functions. 
Tested on the electron bunch trains from the complex multi-gaussian model and 
bunching-enhanced coherent harmonic generation, the algorithm successfully 
reconstructed the temporal signals and achieves better performance than the 
Kramers-Kronig method. This method is expected to crucial provide temporal 
evidence for potential electron beam modulation schemes, and will enable broad 
prospects for future applications.
\end{abstract}

\maketitle


\section{\label{sec:level1}Introduction}

Ultrafast electron bunch trains with femtosecond-scale temporal structures have 
garnered significant interest for diverse applications, including X-ray free-electron 
lasers (XFELs) \cite{Yu2000,Madey1973} and ultrafast electron diffraction for probing atomic and 
molecular ultrafast dynamics \cite{Morimoto2018,Nabben2023}. 
To optimize these applications, accurate 
characterization of the temporal distribution of electron bunch trains is 
necessary, including the pulse spacing, single pulse width and
beam length.

At present, the techniques of longitudinal diagnosis for electron beams
are focused on time-domain and frequency-domain methods \cite{Downer2018}. 
The conventional time-domain methods represented by transverse deflecting 
structures (TDS) can provide direct evidence of the longitudinal profile 
of electron beams but fundamentally limited by the magnitude of the 
transverse deflecting voltage \cite{Alesini2009,Akre2001}, rendering 
them inadequate for requiring femtosecond or sub-femtosecond resolution. 
In contrast, frequency-domain methods have demonstrated more promising 
resolution capabilities and have been widely validated in beam diagnostic 
systems \cite{Castellano2001,Kung1994,Billinghurst2016}. When an electron 
bunch traverses an interface between media with different dielectric 
indices, the coherent/incoherent transition radiation (TR) 
\cite{Frank1966,Schroeder2004} is generated with the angular distribution 
of the radiation spectrum given by 
\begin{equation}
\frac{d^2 W}{d\omega d\Omega}=[N+N^2F(\omega,\theta)]\frac{d^2 W_s}{d\omega d\Omega},
\end{equation}
where $d^2 W_s/(d\omega d\Omega)$ represents the angular distribution of the 
single-electron radiation spectrum. The coherent radiation, which is 
associated with the bunch form factor $F(\omega,\theta)$, enables the 
diagnosis of the longitudinal distribution of electron beams \cite{Lundh2013,Heigoldt2015,LaBerge2024}. 
However, the measured spectrum contains only intensity information, thereby the 
reconstruction of the temporal characteristics constitute the 
critical challenge in frequency-domain diagnostic schemes \cite{Downer2018}.

Although mainstream retrieval algorithms, such as the Kramers-Kronig (KK) 
method \cite{Lai1995,Lai1997} and heuristic iterative methods 
\cite{Bajlekov2013,Heigoldt2015,Bakkali2016}, have been extensively 
developed and rigorously validated for diagnostics of beam profiles with 
simple structures such as single-peak and double-peak configurations, 
their reconstruction capabilities become severely unstable when 
confronted with ultrafast electron bunch trains exhibiting 
complex temporal structures \cite{Zarini2018}, particularly for the signals containing modulations 
across both high- and low-frequency regimes. Fortunately, despite their complex temporal 
structures, ultrafast electron bunch trains still maintain identifiable 
regularities \cite{Lundh2013}. The modulated temporal features correspond analogously 
to distinct spectral modulation signatures, and this intrinsic time-frequency 
correspondence can facilitate the longitudinal diagnostics of bunched electron beams.

In this letter, we present a temporal reconstruction method for 
frequency-domain diagnostics of electron beams with complex microbunching structures. 
Through analysing the intrinsic time-frequency characteristics of the bunch trains, 
the algorithm employs an iterative strategy based on genetic algorithms to accurately 
reconstruct complex temporal modulation profiles. This letter presents a novel 
sampling-based perspective for analyzing coherent transition radiation (CTR) from 
modulated electron bunches, providing a theoretical interpretation of the resulting 
modulations in the spectra. Subsequently, a temporal retrieval algorithm is proposed 
based on the time-frequency relationship, and an adaptive-kernel method is incorporated 
to complete the algorithmic framework, referred to as the adaptive-kernel 
multi-envelope genetic algorithm (AKMEGA). Finally, the proposed algorithm is 
benchmarked against the KK method using the complex modulation cases 
from a multi-Gaussian model and a bunch-enhanced coherent harmonic generation (BECHG) \cite{Feng2024}. 
Owing to its capability to precisely reconstruct the temporal characteristics of 
modulated electron bunches, AKMEGA demonstrates strong competitiveness in 
the field of longitudinal diagnostics of electron bunch trains via spectral analysis.

\section{Temporal-Spectral Properties of Electron Bunch Trains}
To construct the iterative strategy for the proposed algorithm, we first analyze the 
temporal and spectral characteristics of electron bunch trains. While the spectral 
features of CTR signals from electron bunch trains have been previously reported \cite{Lundh2013}, 
the associated theoretical foundations have not yet been discussed. 
To bridge this gap, we present a sampling-based framework that offers a 
rigorous interpretation of the underlying modulation physics.

Notably, temporal modulations in the electron beam are invariably accompanied by 
corresponding modulations in the CTR spectrum. This correspondence becomes 
comprehensible when the bunch train is regarded as a sampling of the envelope 
signal. In mainstream beam modulation schemes, such as coherent harmonic 
generation (CHG), the combination of laser modulators and dispersive sections 
converts energy modulation into density modulation, ultimately yielding a modulated 
electron bunch train with a fixed period corresponding to the laser wavelength 
(or its multiples). In this process, the sampling frequency $\omega_s$ 
corresponds to the laser frequency, which periodically samples the envelope 
signal $e_{nv}(t)$ with temporal intervals $T_s$. Consequently, the longitudinal 
bunch profile of the modulated electron train can be decomposed into a superposition 
of the envelope and the sampling signal:

\begin{equation}
f(t)=e_{nv}(t)s_{amp}(t,T_s).
\end{equation}

The ideal sampling function can be modeled as an infinite sequence of impulse 
with period $T_s$, expressed as $s_{amp}(t)=\sum_n\delta(t-nT_s)$. 
Consequently, the bunch profile can be further elaborated as
\begin{eqnarray}
  f(t) &=& e_{nv}(t)\sum_n\delta(t-nT_s)=\sum_n e_{nv}(t)\delta(t-nT_s) \nonumber\\ 
  &=& \sum_n e_{nv}(nT_s)\delta(t-nT_s),
\end{eqnarray}
which implies that the bunching function assumes the value of the envelope function at 
each temporal sampling point and vanishes elsewhere. Such an idealization presupposes 
the absence of noise and DC components in the electron train, enabling the bunching 
factor to achieve optimal performance at high-order harmonics. However, practical 
sampling functions typically deviate from this idealization, exhibiting finite 
single-pulse broadening, which can be expressed as
\begin{equation}
s_{amp}(t,T_s)=g_s(t,\sigma_t)*\sum_n \delta(t-nT_s),
\end{equation}
where $*$ denotes the convolution operation, with the convolution kernel of a 
Gaussian function $g_s(t,\sigma_t)$. This indicates that the impulse function at 
each sampling point acquires a Gaussian broadening with width $\sigma_t$. 
After applying this sampling function to the envelope, we have
\begin{eqnarray}
  f(t) &=& e_{nv}(t)\times\left[g_s(t,\sigma_t)*\sum_n\delta(t-nT_s)\right]\nonumber\\
       &=& e_{nv}(t)\sum_n g_s(t-nT_s,\sigma_t), \label{eq:5}
\end{eqnarray}
The Gaussian sampling sequence $\sum_n g_s(t-nT_s,\sigma_t)$ samples the envelope 
function $e_{nv}(t)$, with the sampled value of $e_{nv}(nT_s)$ around each 
sampling point accompanied by Gaussian broadening $\sigma_t$. In the 
limit $\sigma_t\to 0$, this model reduces to ideal sampling. 
Conversely, when $\sigma_t\geq T_s$, the Gaussian sampling functions overlap 
and give rise to a DC component in the time domain, precisely matching the 
background DC noise generated by undermodulated cases.

\begin{figure}[htbp]
\includegraphics[width=\linewidth]{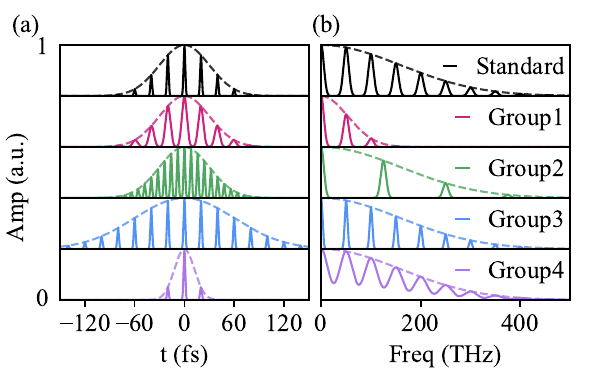}
\caption{\label{fig:1}(a) Temporal profiles and (b) corresponding spectra of electron 
bunch trains with various parameter sets. Compared to the Standard configuration, 
Groups 1-3 exhibit systematic variations in $\sigma_t$, $T_s$ and $B_t$, respectively. 
Group 4 illustrates the spectral overlap induced by broadening when 
$B_{env}>\Omega_s/2$, resulting in the violation of the Nyquist sampling theorem.}
\end{figure}

As for the spectral analysis, the bunching factor $F(\omega)$ is obtained 
via the Fourier transform of $f(t)$, i.e., 
$F(\omega)=\mathcal{F}\left[f(t)\right]=\int f(t)e^{-i\omega t}dt$. Applying 
Fourier transform to the sampling model yields
\begin{eqnarray}
  F(\omega)=\mathcal{F}\left[e_{nv}(t)\times\left[g_s(t,\sigma_t)*\sum_n\delta(t-nT_s)\right]\right]\nonumber\\
  =E_{nv}(\omega)*\left[G_s(\omega,\sigma_\omega)\times\Omega_s\sum_n \delta(\omega-n\Omega_s)\right], \label{eq:6}
\end{eqnarray}
where $\Omega_s=2\pi/T_s$, $E_{nv}(\omega)$ and $G_s(\omega,\sigma_\omega)$ 
denote the Fourier transforms of the envelope function $e_{nv}(t)$ and the 
convolution kernel $g_s(t)$, respectively. Inspection of the convolution term 
$G_s(\omega,\sigma_\omega)\times\Omega_s\sum_n\delta(\omega-n\Omega)$ 
in Eq. \ref{eq:6} reveals that this operation constitutes the sampling of a 
frequency-domain Gaussian function by an impulse train. Owing to the convergent 
integral of the temporal Gaussian function with width $\sigma_t$, its Fourier 
component is likewise convergent, with a spectral width 
$\sigma_\omega=2\pi/\sigma_t$. Therefore, the temporal Gaussian $g(t,\sigma_t)$ 
serves as the envelope $G_s(\omega,\sigma_\omega)$ being sampled in the frequency 
domain, and this further implies
\begin{eqnarray}
  F(\omega)&=&E_{nv}*\Omega_s\sum_n G_s(n\Omega_s)\delta(\omega-n\Omega_s)\nonumber \\
  &=&\Omega_s\sum_n G_s(n\Omega_s)E_{nv}(\omega-n\Omega_s), \label{eq:7}
\end{eqnarray}
which illustrates that the temporal envelope $e_{nv}(t)$ manifests in the frequency 
domain as a broadened sampling function. Since physical signals 
are inherently continuous and integrable, $E_{nv}(\omega)$ should be an 
integrable function with finite bandwidth $B_{env}$. Equation \ref{eq:7} can thus 
be interpreted as follows: at each spectral sampling point, a replica of 
$E_{nv}(\omega)$ emerges, characterized by bandwidth $B_{env}$ and amplitude 
$G_s(n\Omega_s)$. When $B_{env}\leq\Omega_s/2$, these replicas remain spectrally 
separated, enabling the reconstruction of the original envelope $e_{nv}(t)$ from 
any individual replica $E_{nv}(\omega-n\Omega)$ in accordance with the Nyquist 
sampling theorem \cite{oppenheim1996sampling}. However, the envelope bandwidth $B_{env}$ typically exceeds the 
sampling frequency in ultrafast electron bunch trains, especially for 
electron beams from laser-wakefield accelerators (LWFAs) with temporal duration below 
30 fs inherently limited by plasma density \cite{Esarey2009}. 
When $B_{env}> \Omega_s/2$, spectral overlap occurs between adjacent 
replicas as shown in Group 4 of Fig. \ref{fig:1}, precluding the signal recovery in time-domain.

A comparison of Eqs. \ref{eq:5} and \ref{eq:7} reveals a explicit duality between 
the sampling of the temporal and spectral domains. An increase 
in the temporal sampling interval $T_s$ results in a decreased spectral 
sampling interval $\Omega_s$. Similarly, a larger bandwidth 
$B_{env}$ corresponds to a shorter bunch total length, whereas a broader 
spectral width $\sigma_\omega$ reflects a sharper single-pulse profile 
$\sigma_t$ in the time domain, concomitantly yielding an enhanced bunching 
factor. As an illustration, Figure \ref{fig:1}(a) presents the temporal bunch profiles 
of electron trains under various parameter sets, all described by
\begin{equation}
  f(t)=\frac{1}{2\pi B_t\sigma_t}\exp\left(-\frac{t^2}{2B_t^2}\right)\sum_n \exp\left[-\frac{\left(t-nT_s\right)^2}{2\sigma_t^2}\right]. \label{eq:8}
\end{equation}
For simplicity, the temporal envelope function $e_{nv}(t)$ is likewise modeled as 
a Gaussian with width parameter $B_t$. Specific parameter values are summarized in 
Table 1. The corresponding spectral representation is obtained via the Fourier 
transform of Eq. \ref{eq:8}
\begin{equation}
  F(\omega)=\Omega_s\sum_n \exp\left(-\frac{n^2\Omega_s^2}{2\sigma_\omega^2}\right)\exp\left[-\frac{\left(\omega-n\Omega_s\right)^2}{2B_\omega^2}\right].
\end{equation}
The corresponding spectra are plotted in Fig. \ref{fig:1}(b) after retaining 
only the positive-frequency components. The duality properties discussed above are 
easily discernible from these spectra, offering valuable insights for 
the development of temporal reconstruction algorithms for CTR spectroscopy of 
electron bunch trains.

\begin{table}[htbp]
\caption{Parameters of electron bunch trains employed for testing. Here, $\sigma_t$ 
denotes the single-pulse width, $T_s$ the modulation period, and $B_t$ the envelope 
width.\label{tab:table1}}
\begin{ruledtabular}
\begin{tabular}{cccccc}
\textrm{Parameter}&
\textrm{Standard}&
\textrm{Group 1}&
\textrm{Group 2}&
\textrm{Group 3}&
\textrm{Group 4}\\
\colrule
$\sigma_t$ & \textrm{1 fs} & \textrm{3 fs} & \textrm{1 fs} & \textrm{1 fs} & \textrm{1 fs}\\
$T_s$ & \textrm{20 fs} & \textrm{20 fs} & \textrm{10 fs} & \textrm{20 fs} & \textrm{20 fs}\\
$B_t$ & \textrm{30 fs} & \textrm{30 fs} & \textrm{30 fs} & \textrm{60 fs} & \textrm{10 fs}\\
\end{tabular}
\end{ruledtabular}
\end{table}

\section{adaptive-kernel multi-envelope genetic algorithm}
\subsection{Multi-envelope Genetic Algorithm}
\begin{figure*}[htb]
\includegraphics[width=\linewidth]{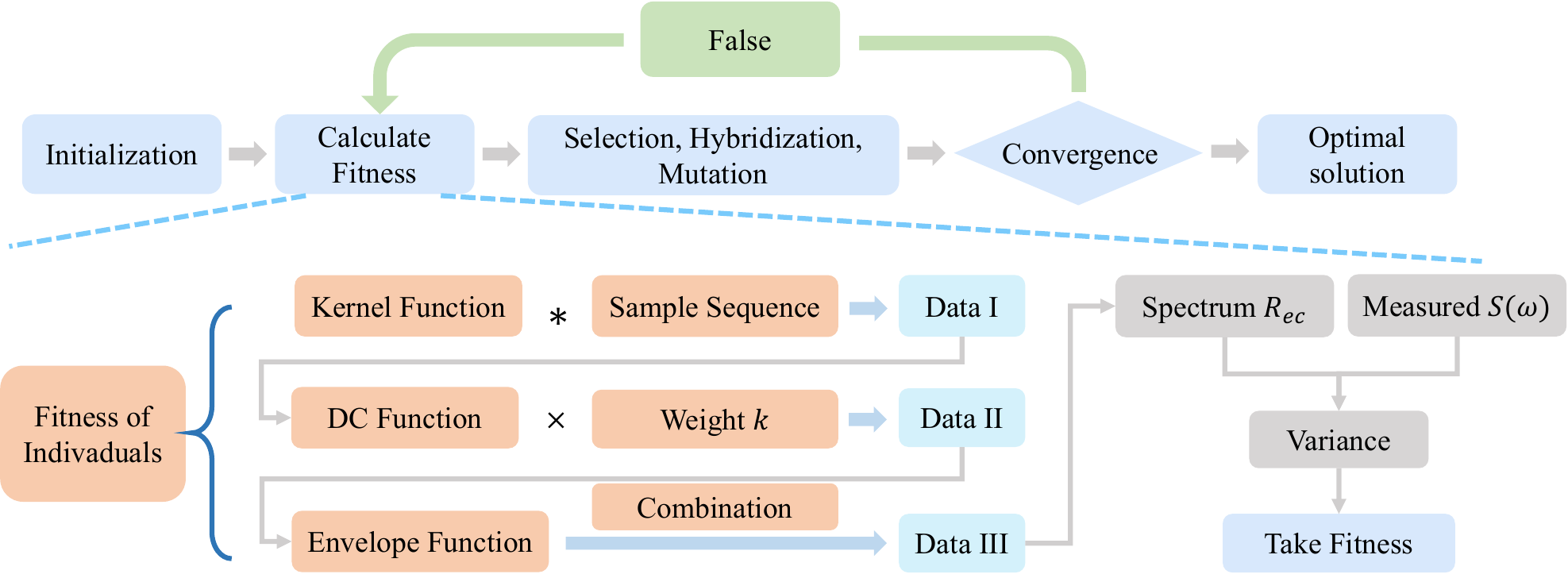}
\caption{\label{fig:2}Flowchart of the multi-envelope genetic algorithm. 
The framework is predicated upon a classical genetic algorithm, 
wherein the high- and low-frequency components are decoupled and reconstructed 
independently, and finally combined to yield the complete temporal solution.}
\end{figure*}

Constrained by low-frequency noise and practical modulation 
performance, the condition $B_{env}\geq \Omega_s/2$ is prevalent across various 
modulation schemes (e.g. CHG). According to the 
Nyquist sampling theorem, spectral overlap in the frequency domain prevents the 
recovery of the envelope function. Fortunately, as elucidated in the preceding 
analysis, decoupling the high-frequency component (sampling) from the 
low-frequency component (envelope) enables the algorithm to impose 
sparsity upon the envelope while retaining the dense information in the 
high-frequency sampling sequence. This enables the adoption of a sparse 
representation strategy analogous to that employed in sparse algorithm, 
wherein the temporal envelope is constructed from a limited set of basis 
functions and subsequently employed as the sampled function to reconstruct 
the complete temporal CTR signal \cite{Shechtman2014,Su2018}. 
Different from the basis of Gaussian functions utilized in most cases, this 
algorithm employs a linear combination of Butterworth functions to accommodate 
more complex scenarios due to the additional degree of freedom provided by their 
adjustable order, which allows for precise control over the shape and steepness of 
the basis functions \cite{1929Butterworth}, expressed as
\begin{equation}
  K(t)=\sum_i\frac{w_i}{1+\left|\frac{t-P_i}{W_i/2}\right|^{N_i}},
\end{equation}
where the i-th Butterworth function is characterized by its amplitude weight $w_i$, 
central position $P_i$, and width $W_i$, while the rise and fall slopes are 
governed by the exponent parameter $N_i$. With four independent parameters per 
basis function, one may readily balance the requisite number of functions against 
computational cost.

For the high-frequency components in the spectra, the sampling function is 
obtained via the convolution of an impulse train with the single-pulse 
kernel function, i.e., $s_{amp}(t)=g_s(t,\sigma_t)*\delta(t-nT_s)$, 
where $g_s(t,\sigma_t)$ denotes the Gaussian kernel. 
It is noteworthy that this scheme inherently assumes uniformity in the 
pulse morphology across the bunch train, as the kernel function is 
required to remain invariant during sampling. 
If the pulses within the modulated train exhibit heterogeneous morphologies, i.e., 
some being broad and others sharp, the algorithm will adopt an averaged profile as 
the convolution kernel and construct the temporal solution. 
Furthermore, the cases of absent modulation can be identified by 
imposing manual judgment criteria to flag reconstruction failures when the genetic 
algorithm fails to converge.

The target temporal function constructed in each iteration is given by
\begin{equation}
  f(t)=\left[s_{amp}(t,T_s)+kK_{DC}(t)\right]\cdot K_{env}(t),
\end{equation}
where $K_{DC}(t)$ and $K_{env}(t)$ denote the Butterworth functions 
fitting the DC component and envelope with parameter $k$ 
regulating the DC fraction. As shown in Fig \ref{fig:2}, the algorithmic framework 
is based on a genetic algorithm, wherein an extensive set of randomized 
initial populations undergoes selection, hybridization and mutation to 
evolve superior solutions. Within each population, the 
recovery temporal profile $f(t)$ is constructed and subsequently 
transformed via FFT, retaining only positive frequencies to compute the 
reconstructed spectrum $Rec(\omega)$. The critical point in each iteration 
is formulating the objective function and further computing the fitness, 
which is defined as
\begin{equation}
  F_{it}=\frac{1}{\left<S(\omega)-Rec(\omega)\right>^2},
\end{equation}
where $S(\omega)$ denotes the measured CTR spectrum and angular brackets indicate 
ensemble averaging. Upon convergence of the fitness through successive 
iterations, the optimal population is selected as the temporal reconstruction 
solution.

\subsection{Adaptive-Kernel Function}

To impose temporal constraints and enhance phase retrieval performance, 
prior algorithms have predominantly relied upon Gaussian assumptions \cite{Su2018,Zarini2018}, 
i.e., individual electron bunches exhibit Gaussian or 
quasi-Gaussian distributions, and the benchmarks employed in these studies were 
also formulated with Gaussian assumptions. 
However, in most practical scenarios, presuming quasi-Gaussian pulse profiles is 
typically excessive, particularly for complex modulation schemes where individual 
pulses cannot be adequately described by a single Gaussian function. 
Building upon preceding signal analysis of electron bunch trains, we 
propose a framework for extracting the single-pulse profile as 
the kernel function for convolution. In contrast to Gaussian assumptions, 
the adaptive kernel enables a more faithful reconstruction of the electron 
pulse morphology directly from the measured spectrum.

Due to the duality between temporal and spectral sampling 
as discussed before, the spectral envelope inherently encodes information 
regarding the temporal single-pulse morphology. The concept of the 
adaptive-kernel approach lies in extracting spectral envelope and subsequently 
reconstructing the temporal kernel function. In most laser-modulating 
schemes such as CHG, the modulation frequency associated with the vicinity 
of the laser frequency or its integer harmonics is typically available as 
prior parameter. This enables the implementation of a variable search for the 
sampling frequency, with a reasonable search range defined around the laser 
frequency. The measured spectrum is then sampled at this frequency and fitted 
with a spline interpolation to obtain the envelope. Temporal retrieval of single pulse 
is subsequently performed on the fitted spectral envelope by using the KK algorithm, 
yielding the numerical solution for the temporal convolution kernel.

\begin{figure}[htb]
\includegraphics[width=\linewidth]{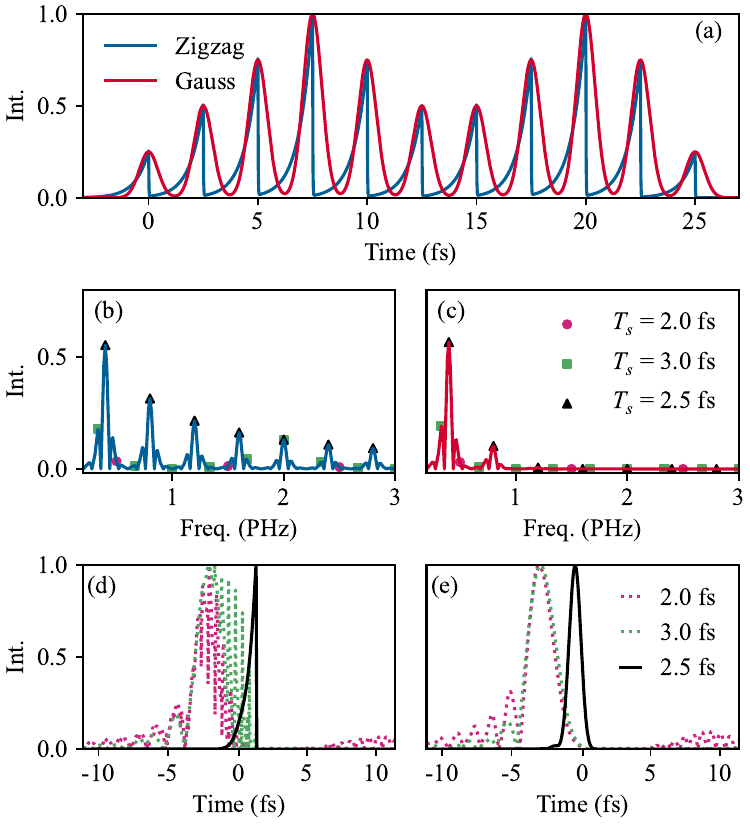}
\caption{\label{fig:3}Reconstruction of the temporal convolution kernel. (a) 
Temporal profiles of Zigzag (blue) and standard Gaussian model (red). 
Spectra of the (b) Zigzag signal and (c) Gaussian signal, with sampling points 
scattered in plots. Kernel functions reconstructed via the KK algorithm are shown 
in (d) and (e) correspondingly, alongside the reconstructed functions 
with a $\pm0.5$ fs sampling interval offset (dashed curves).}
\end{figure}

To further illustrate the extraction of the temporal convolution kernel, we 
configured two benchmark signals featuring distinct single-pulse profiles, as 
shown in Fig. \ref{fig:3}(a). Both signals consist of several pulses with a 
spacing of $\sigma_i=2.5$ fs and random amplitude weights $\alpha_i$. The first benchmark 
employs zigzag pulses with a width of 1 fs, characterized by a gradual rising 
edge and a steep falling edge, while the second utilizes standard Gaussian functions 
for comparison. Figures \ref{fig:3}(b) and \ref{fig:3}(c) present the spectral profiles of these two 
signals, where the sampling points used for reconstructing the convolution kernels 
are indicated. The reconstructed kernel functions obtained via the KK 
algorithm are plotted in Figs. \ref{fig:3}(d) and \ref{fig:3}(e). For comparison, we also examined 
the scenario involving a shift of $\pm0.5$ fs in the sampling interval $T_s$. 
The corresponding sampling points are marked in the figures, and the reconstructed 
kernels are depicted as dashed curves.

In the present algorithm, the construction of the adaptive kernel is not 
predicated upon any specific functional form, thereby obviating the need for 
prior assumptions. The sampling frequency upon which this method relies is 
generally available as a priori information in most electron beam modulation 
schemes, permitting a reasonable estimation of the search range. When the search 
variable deviates from the true sampling frequency, the constructed adaptive 
kernel exhibits pronounced oscillations, as illustrated by the dashed 
curves in Figs. \ref{fig:3}(d) and \ref{fig:3}(e). This, in turn, causes the reconstructed 
spectral envelope to deviate from the measured spectrum, leading to the 
elimination of such populations during the evolution in genetic algorithm. 
Employing the adaptive kernel as the convolution kernel for constructing 
the complete temporal signal substantially enhances both reconstruction 
efficiency and fidelity.

\section{Reconstruction of Highly Modulated Bunch Trains}
\subsection{Reconstruction of Complex Multi-Gaussian Model Bunch}

\begin{figure}[htb]
\includegraphics[width=\linewidth]{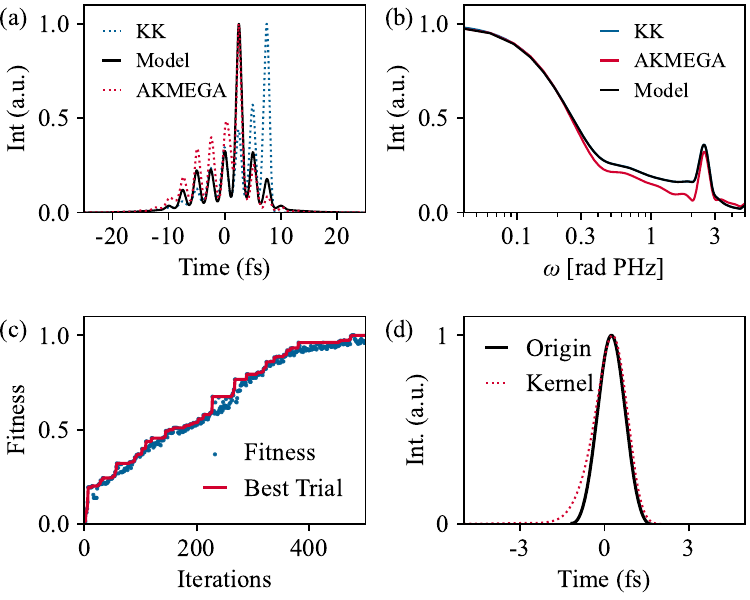}
\caption{\label{fig:4}Reconstruction of the complex multi-Gaussian model bunch. 
(a) The retrieval temporal signals and (b) the spectral profiles, 
presenting the results of AKMEGA and the KK algorithm alongside 
the original signal, with the spectral data scattered in the figure for 
the reconstruction. (c) The convergence process of AKMEGA. (d) The 
adaptive kernel function recovered via inverse sampling (solid) and the model single pulse 
near the origin for comparison (dash).}
\end{figure}

In this case, a pedestal Gaussian with rms width $\sigma_1=20$ fs is modulated by 9 
Gaussians, each with $\sigma_i=0.5$ fs, separated by 2.5 fs and weighted with 
$\alpha_i=0.1, 0.5, 1.0, 1.0, 1.5, 5.0, 1.5, 0.8$ and $0.1$, respectively. 
Additionally, a Gaussian DC component with a width of 8 fs weighted with 0.2 is 
superimposed. The temporal profile of this multi-gaussian Model is shown in Fig. \ref{fig:4}(a), 
and the original as well as reconstructed spectra are presented in Fig. \ref{fig:4}(b). 
According to the spectral response range of the detector, only the components 
above 40 rad THz are employed for the algorithm.

The genetic algorithm was configured with 10,000 random initial points 
and a maximum of 500 iterations. The search space for the position and width of 
the Butterworth functions was constrained within a 100 fs window. 
Similarly, the search for the sampling frequency, required to construct the 
temporal convolution kernel, was limited to the range of 2-3 fs. 
The convergence process is illustrated in Fig. \ref{fig:4}(c), 
showing that the algorithm converged after approximately 400 generations to 
yield a reasonable reconstruction, and the optimal population was selected as the 
reconstruction solution. Results are presented in Fig. \ref{fig:4}(a) in comparison 
with the original signal, and the coefficient of determination, $R^2=0.87$, 
indicates that the original temporal signal was recovered with high accuracy.

For comparison, the reconstruction results obtained via the KK algorithm are also 
presented in Fig. \ref{fig:4}(a). Although the KK algorithm provides a roughly similar 
sketch to the original signal, discrepancies in local details prevent an 
exact match. This deviation arises from the non-negligiblity of the Blaschke phase, 
which has been analyzed in other studies \cite{Lai1995}. Specifically, when weak components 
precede strong ones, the Blaschke phase significantly influences the 
reconstruction, and the invalid minimum phase assumption leads to substantial 
deviations between the reconstructed bunch shape and the actual profile.

\subsection{Reconstruction of Bunching-Enhanced Coherent Harmonic Generation}

\begin{figure}[htb]
\includegraphics[width=\linewidth]{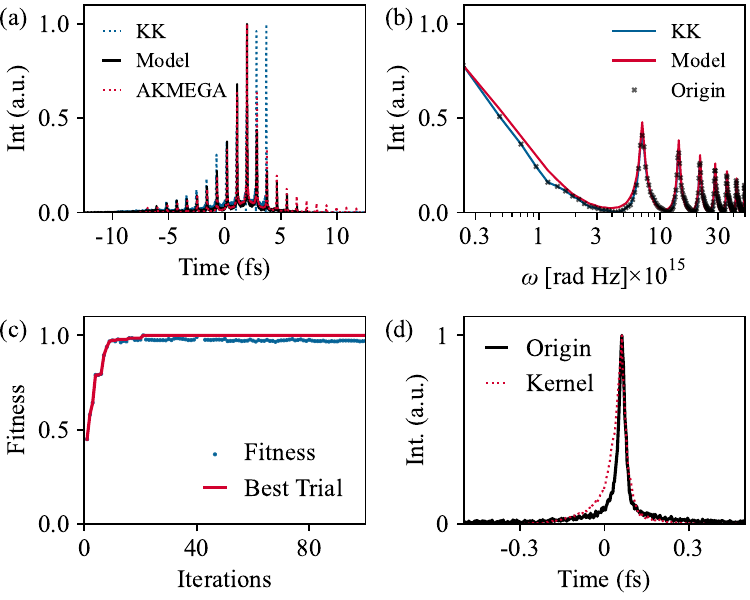}
\caption{\label{fig:5}Reconstruction of the bunch train from BECHG. 
(a) The retrieval temporal signals and (b) the spectral profiles, 
presenting the results of AKMEGA and the KK algorithm alongside 
the original signal, with the spectral data scattered in the figure for 
the reconstruction. (c) The convergence process of AKMEGA. (d) The 
adaptive kernel function recovered via inverse sampling (solid) and the model single pulse 
near the origin for comparison (dash).}
\end{figure}

Bunching-enhanced coherent harmonic generation (BECHG), as a prebunching scheme for 
LWFA electron beams to achieve an ultra-high bunching factor, has received 
considerable attention \cite{Feng2024}. As a challenging case for the algorithms, 
the signal employed is derived from the bunching profile in \cite{Feng2024}. Since the 
third harmonic of a Ti:sapphire laser was used to modulate the electron 
beam, the modulation frequency is centered around 266 nm. The temporal profile of 
the electron bunch train is displayed in Fig. \ref{fig:5}(a), while the corresponding 
spectrum is shown in Fig. \ref{fig:5}(b). In addition to the high-frequency modulation, 
the signal also contains a DC component with relatively weak intensity.

This complex configuration would likely encounter issues such as solution ambiguity 
and mismatch in most algorithms. However, the proposed algorithm further recovers 
the complete temporal signal after reconstructing the kernel function. The adaptive 
kernel function derived from the sampled reconstruction is presented in Fig. \ref{fig:5}(d), 
nearby the actual single-peak signal closed to $t=0$ for comparison.

Based on the reconstructed kernel function, the complete temporal signal was 
obtained by convolving the kernel with an impulse train, and subsequently 
combined with Butterworth envelopes and DC components to match the measured 
spectrum. Starting from 10,000 random initial points, the genetic algorithm finally  
converged within 100 iterations, as shown in Fig. \ref{fig:5}(c). The reconstruction 
results are presented in Fig. \ref{fig:5}(a), 
while the reconstructed spectrum in Fig. \ref{fig:5}(b). A coefficient of determination 
$R^2$ of 0.74 indicates that the longitudinal reconstruction of the electron 
bunch train was essentially achieved. For comparison, the results from the KK 
algorithm are also plotted. Since this case involves at least 
one zero in the complex frequency plane, the KK algorithm almost failed to 
reconstruct this temporal signal.

\section{Conclusion}
In summary, we have proposed a temporal reconstruction algorithm tailored for complex 
modulated electron beams. Based on the signal analysis of modulated electron 
bunch trains, this method enables spectral diagnostics across various 
complex modulation scenarios. We first analyze the relationship between the 
time and frequency domain of the modulated signal and provided an interpretation 
of the time-frequency duality from a sampling perspective. Subsequently, we 
separates and reconstructs the high- and low-frequency components. For the 
high-frequency reconstruction, the algorithm reconstructs a single-peak function 
to serve as a convolution kernel by using the prior knowledge of 
the modulation period inherent to the specific schemes. For the low-frequency 
reconstruction, the sparsity of the separated envelope and DC component enable 
recovery by employing several Butterworth basis functions to reproduce the temporal 
envelope. Finally, the algorithm combines the high- and low-frequency components, 
calculating the reconstructed spectrum and minimizing the residuals with respect 
to the measured spectrum. The advantage of AKMEGA lies in its ability to 
reconstruct signals in two different scales without the reliance on Gaussian 
assumptions. It successfully addresses the challenge of temporal recovery in 
spectral diagnostics for complex modulation schemes, providing a critical 
diagnostic solution for applications such as X-ray free electron lasers.

\section*{acknowledgments}

This work was supported by the National Natural Science Foundation of China 
(Grant No. 12388102), the Strategic Priority Research Program of the Chinese Academy 
of Sciences (Grant No. XDB0890200), the National Natural Science Foundation of China 
(Grant Nos. 12225411 and 12474349), the National Key R\&D Program of China (Grant No. 
2025YFF0515000), the CAS Project for Young Scientists in Basic Research (Grant No. 
YSBR060), the CAS Youth Innovation Promotion Association (No. 2022242), the Ningbo 
Yongjiang Talent Programme (Grant No. 2025A-104-G), Zhangjiang Laboratory, and the 
New Cornerstone Science Foundation through the XPLORER PRIZE.

\section{Data Availability}
The data that support the findings of this article are available from the corresponding author upon the reasonable request.

\nocite{*}

\bibliography{AKMEGA}

@ARTICLE{Morimoto2018,
  author  = {Y. Morimoto and P. Baum},
  title   = {Diffraction and microscopy with attosecond electron pulse trains},
  journal = {Nature Physics},
  year    = {2018},
  volume  = {14},
  number  = {3},
  pages   = {252--256},
  month   = mar,
  doi     = {10.1038/s41567-017-0007-6},
  url     = {https://doi.org/10.1038/s41567-017-0007-6},
}

@ARTICLE{Nabben2023,
  author  = {D. Nabben and J. Kuttruff and L. Stolz and A. Ryabov and P. Baum},
  title   = {Attosecond electron microscopy of sub-cycle optical dynamics},
  journal = {Nature},
  year    = {2023},
  volume  = {619},
  number  = {7968},
  pages   = {63--67},
  month   = jul,
  doi     = {10.1038/s41586-023-06074-9},
}

@ARTICLE{Madey1973,
  author={Madey, J. M. J. and Schwettman, H. A. and Fairbank, W. M.},
  journal={IEEE Transactions on Nuclear Science}, 
  title={A Free Electron Laser}, 
  year={1973},
  volume={20},
  number={3},
  pages={980-983},
  doi={10.1109/TNS.1973.4327304}}

@article{Yu2000,
author = {L.-H. Yu  and M. Babzien  and I. Ben-Zvi  and L. F. DiMauro  and A. Doyuran  and W. Graves  and E. Johnson  and S. Krinsky  and R. Malone  and I. Pogorelsky  and J. Skaritka  and G. Rakowsky  and L. Solomon  and X. J. Wang  and M. Woodle  and V. Yakimenko  and S. G. Biedron  and J. N. Galayda  and E. Gluskin  and J. Jagger  and V. Sajaev  and I. Vasserman },
title = {High-Gain Harmonic-Generation Free-Electron Laser},
journal = {Science},
volume = {289},
number = {5481},
pages = {932-934},
year = {2000},
doi = {10.1126/science.289.5481.932},
URL = {https://www.science.org/doi/abs/10.1126/science.289.5481.932}}

@article{Downer2018,
  title = {Diagnostics for plasma-based electron accelerators},
  author = {Downer, M. C. and Zgadzaj, R. and Debus, A. and Schramm, U. and Kaluza, M. C.},
  journal = {Rev. Mod. Phys.},
  volume = {90},
  issue = {3},
  pages = {035002},
  numpages = {62},
  year = {2018},
  month = {Aug},
  publisher = {American Physical Society},
  doi = {10.1103/RevModPhys.90.035002},
  url = {https://link.aps.org/doi/10.1103/RevModPhys.90.035002}
}

@article{Alesini2009,
  title = {rf deflector design of the CLIC test facility CTF3 delay loop and beam loading effect analysis},
  author = {Alesini, David and Marcellini, Fabio},
  journal = {Phys. Rev. ST Accel. Beams},
  volume = {12},
  issue = {3},
  pages = {031301},
  numpages = {16},
  year = {2009},
  month = {Mar},
  publisher = {American Physical Society},
  doi = {10.1103/PhysRevSTAB.12.031301},
  url = {https://link.aps.org/doi/10.1103/PhysRevSTAB.12.031301}
}

@INPROCEEDINGS{Akre2001,
  author={Akre, R. and Bentson, L. and Emma, P. and Krejcik, P.},
  booktitle={PACS2001. Proceedings of the 2001 Particle Accelerator Conference (Cat. No.01CH37268)}, 
  title={A transverse rf deflecting structure for bunch length and phase space diagnostics}, 
  year={2001},
  volume={3},
  number={},
  pages={2353-2355 vol.3},
  doi={10.1109/PAC.2001.987379}}

@article{Castellano2001,
  title = {Measurements of coherent diffraction radiation and its application for bunch length diagnostics in particle accelerators},
  author = {Castellano, M. and Verzilov, V. A. and Catani, L. and Cianchi, A. and Orlandi, G. and Geitz, M.},
  journal = {Phys. Rev. E},
  volume = {63},
  issue = {5},
  pages = {056501},
  numpages = {8},
  year = {2001},
  month = {Apr},
  publisher = {American Physical Society},
  doi = {10.1103/PhysRevE.63.056501},
  url = {https://link.aps.org/doi/10.1103/PhysRevE.63.056501}
}

@article{Kung1994,
  title = {Generation and Measurement of 50-fs (rms) Electron Pulses},
  author = {Kung, Pamela and Lihn, Hung-chi and Wiedemann, Helmut and Bocek, David},
  journal = {Phys. Rev. Lett.},
  volume = {73},
  issue = {7},
  pages = {967--970},
  numpages = {0},
  year = {1994},
  month = {Aug},
  publisher = {American Physical Society},
  doi = {10.1103/PhysRevLett.73.967},
  url = {https://link.aps.org/doi/10.1103/PhysRevLett.73.967}
}

@article{Billinghurst2016,
  title = {Longitudinal bunch dynamics study with coherent synchrotron radiation},
  author = {Billinghurst, B. E. and Bergstrom, J. C. and Baribeau, C. and Batten, T. and May, T. E. and Vogt, J. M. and Wurtz, W. A.},
  journal = {Phys. Rev. Accel. Beams},
  volume = {19},
  issue = {2},
  pages = {020704},
  numpages = {11},
  year = {2016},
  month = {Feb},
  publisher = {American Physical Society},
  doi = {10.1103/PhysRevAccelBeams.19.020704},
  url = {https://link.aps.org/doi/10.1103/PhysRevAccelBeams.19.020704}
}

@article{Frank1966,
doi = {10.1070/PU1966v008n05ABEH003034},
url = {https://doi.org/10.1070/PU1966v008n05ABEH003034},
year = {1966},
month = {may},
publisher = {},
volume = {8},
number = {5},
pages = {729},
author = {I M Frank},
title = {TRANSITION RADIATION AND OPTICAL PROPERTIES OF MATTER},
journal = {Soviet Physics Uspekhi}
}

@article{Schroeder2004,
  title = {Theory of coherent transition radiation generated at a plasma-vacuum interface},
  author = {Schroeder, C. B. and Esarey, E. and van Tilborg, J. and Leemans, W. P.},
  journal = {Phys. Rev. E},
  volume = {69},
  issue = {1},
  pages = {016501},
  numpages = {12},
  year = {2004},
  month = {Jan},
  publisher = {American Physical Society},
  doi = {10.1103/PhysRevE.69.016501},
  url = {https://link.aps.org/doi/10.1103/PhysRevE.69.016501}
}

@article{Lundh2013,
  title = {Experimental Measurements of Electron-Bunch Trains in a Laser-Plasma Accelerator},
  author = {Lundh, O. and Rechatin, C. and Lim, J. and Malka, V. and Faure, J.},
  journal = {Phys. Rev. Lett.},
  volume = {110},
  issue = {6},
  pages = {065005},
  numpages = {5},
  year = {2013},
  month = {Feb},
  publisher = {American Physical Society},
  doi = {10.1103/PhysRevLett.110.065005},
  url = {https://link.aps.org/doi/10.1103/PhysRevLett.110.065005}
}

@article{Heigoldt2015,
  title = {Temporal evolution of longitudinal bunch profile in a laser wakefield accelerator},
  author = {Heigoldt, M. and Popp, A. and Khrennikov, K. and Wenz, J. and Chou, S. W. and Karsch, S. and Bajlekov, S. I. and Hooker, S. M. and Schmidt, B.},
  journal = {Phys. Rev. ST Accel. Beams},
  volume = {18},
  issue = {12},
  pages = {121302},
  numpages = {6},
  year = {2015},
  month = {Dec},
  publisher = {American Physical Society},
  doi = {10.1103/PhysRevSTAB.18.121302},
  url = {https://link.aps.org/doi/10.1103/PhysRevSTAB.18.121302}
}

@ARTICLE{LaBerge2024,
  author  = {M. LaBerge and B. Bowers and Y.-Y. Chang and J. Couperus Cabada{\`{g}} and A. Debus and A. Hannasch and R. Pausch and S. Sch{\"{o}}bel and J. Tiebel and P. Ufer and A. Willmann and O. Zarini and R. Zgadzaj and A. H. Lumpkin and U. Schramm and A. Irman and M. C. Downer},
  title   = {Revealing the three-dimensional structure of microbunched plasma-wakefield-accelerated electron beams},
  journal = {Nature Photonics},
  year    = {2024},
  volume  = {18},
  number  = {9},
  pages   = {952--959},
  month   = sep,
}

@article{Lai1997,
title = {On using the coherent far IR radiation produced by a charged-particle bunch to determine its shape: I Analysis},
journal = {Nuclear Instruments and Methods in Physics Research Section A: Accelerators, Spectrometers, Detectors and Associated Equipment},
volume = {397},
number = {2},
pages = {221-231},
year = {1997},
issn = {0168-9002},
doi = {https://doi.org/10.1016/S0168-9002(97)00690-6},
url = {https://www.sciencedirect.com/science/article/pii/S0168900297006906},
author = {R Lai and A.J Sievers}
}

@article{Lai1995,
  title = {Phase problem associated with the determination of the longitudinal shape of a charged particle bunch from its coherent far-ir spectrum},
  author = {Lai, R. and Sievers, A. J.},
  journal = {Phys. Rev. E},
  volume = {52},
  issue = {4},
  pages = {4576--4579},
  numpages = {0},
  year = {1995},
  month = {Oct},
  publisher = {American Physical Society},
  doi = {10.1103/PhysRevE.52.4576},
  url = {https://link.aps.org/doi/10.1103/PhysRevE.52.4576}
}

@article{Bajlekov2013,
  title = {Longitudinal electron bunch profile reconstruction by performing phase retrieval on coherent transition radiation spectra},
  author = {Bajlekov, S. I. and Heigoldt, M. and Popp, A. and Wenz, J. and Khrennikov, K. and Karsch, S. and Hooker, S. M.},
  journal = {Phys. Rev. ST Accel. Beams},
  volume = {16},
  issue = {4},
  pages = {040701},
  numpages = {14},
  year = {2013},
  month = {Apr},
  publisher = {American Physical Society},
  doi = {10.1103/PhysRevSTAB.16.040701},
  url = {https://link.aps.org/doi/10.1103/PhysRevSTAB.16.040701}
}

@article{Bakkali2016,
  title = {Electron bunch profile reconstruction based on phase-constrained iterative algorithm},
  author = {Bakkali Taheri, F. and Konoplev, I. V. and Doucas, G. and Baddoo, P. and Bartolini, R. and Cowley, J. and Hooker, S. M.},
  journal = {Phys. Rev. Accel. Beams},
  volume = {19},
  issue = {3},
  pages = {032801},
  numpages = {7},
  year = {2016},
  month = {Mar},
  publisher = {American Physical Society},
  doi = {10.1103/PhysRevAccelBeams.19.032801},
  url = {https://link.aps.org/doi/10.1103/PhysRevAccelBeams.19.032801}
}

@PROCEEDINGS{Zarini2018,
  author={Zarini, Omid and Köhler, Alexander and Couperus, Jurjen and Pausch, Richard and Kurz, Thomas and Schöbel, Susanne and Meißner, Heide and Bussmann, Michael and Schramm, Ulrich and Irman, Arie and Debus, Alexander},
  booktitle={2018 IEEE Advanced Accelerator Concepts Workshop (AAC)}, 
  title={Advanced Methods for Temporal Reconstruction of Modulated Electron Bunches}, 
  year={2018},
  volume={},
  number={},
  pages={1-5},
  doi={10.1109/AAC.2018.8659388}
}

@article{Feng2024,
    author = {Feng, Ke and Jiang, Kangnan and Hu, Runshu and Luan, Shixia and Wang, Wentao and Li, Ruxin},
    title = {Bunching enhancement for coherent harmonic generation by using phase merging effects},
    journal = {Matter and Radiation at Extremes},
    volume = {9},
    number = {5},
    pages = {057201},
    year = {2024},
    month = {07},
    issn = {2468-2047},
    doi = {10.1063/5.0191508},
    url = {https://doi.org/10.1063/5.0191508}
}

@incollection{oppenheim1996sampling,
  author    = {Alan V. Oppenheim and Alan S. Willsky and S. Hamid Nawab},
  title     = {Sampling},
  booktitle = {Signals and Systems},
  chapter   = {8},
  pages     = {514--521},
  year      = {1996},
  edition   = {2nd},
  publisher = {Prentice Hall},
  address   = {Upper Saddle River, NJ}
}

@article{Esarey2009,
  title = {Physics of laser-driven plasma-based electron accelerators},
  author = {Esarey, E. and Schroeder, C. B. and Leemans, W. P.},
  journal = {Rev. Mod. Phys.},
  volume = {81},
  issue = {3},
  pages = {1229--1285},
  numpages = {0},
  year = {2009},
  month = {Aug},
  publisher = {American Physical Society},
  doi = {10.1103/RevModPhys.81.1229},
  url = {https://link.aps.org/doi/10.1103/RevModPhys.81.1229}
}

@article{1929Butterworth,
  title={On the theory of filter amplifiers},
  author={Butterworth, S. },
  journal={wireless engineer},
  year={1929},
}

@ARTICLE{Shechtman2014,
  author={Shechtman, Yoav and Beck, Amir and Eldar, Yonina C.},
  journal={IEEE Transactions on Signal Processing}, 
  title={GESPAR: Efficient Phase Retrieval of Sparse Signals}, 
  year={2014},
  volume={62},
  number={4},
  pages={928-938},
  doi={10.1109/TSP.2013.2297687}}

@article{Su2018,
  title = {Temporal diagnostics of femtosecond electron bunches with complex structures using sparsity-based algorithm},
  author = {Su, Q. Q. and Hua, J. F. and Nie, Z. and Ma, Y. and Liu, S. and Zheng, Y. F. and Pai, C.-H. and Lu, W.},
  journal = {Phys. Rev. Accel. Beams},
  volume = {21},
  issue = {11},
  pages = {112801},
  numpages = {7},
  year = {2018},
  month = {Nov},
  publisher = {American Physical Society},
  doi = {10.1103/PhysRevAccelBeams.21.112801},
  url = {https://link.aps.org/doi/10.1103/PhysRevAccelBeams.21.112801}
}

@CONTROL{REVTEX42Control}

@CONTROL{apsrev42Control,author="08",editor="1",pages="0",title="0",year="1"}

\end{document}